\documentstyle[prl,eqsecnum,aps]{revtex}

\begin{document}
\author{Jian-Qi Shen $^{1,}$$^{2}$ \footnote{E-mail address: jqshen@coer.zju.edu.cn}}
\address{$^{1}$  Centre for Optical
and Electromagnetic Research, State Key Laboratory of Modern
Optical Instrumentation, \\Zhejiang University,
Hangzhou Yuquan 310027, P.R. China\\
$^{2}$ Zhejiang Institute of Modern Physics and Department of
Physics, Zhejiang University, Hangzhou 310027, P.R. China}
\date{\today}
\title{Three kinds of compact thin subwavelength cavity resonators \\containing left-handed media: rectangular, cylindrical, spherical}
\maketitle

\begin{abstract}
In the present paper we investigate the restriction conditions for
three kinds of cavity resonators ({\it i.e.}, the rectangular,
cylindrical, spherical resonators). It is shown that the layer of
materials with negative optical refractive indices can act as a
phase compensator/conjugator, and thus by combining such a layer
with another layer made of the regular medium one can obtain a
so-called compact thin subwavelength cavity resonator.

\end{abstract}
\section{Introduction}
More recently, a kind of artificial composite metamaterials (the
so-called {\it left-handed media}) having a frequency band where
the effective permittivity and the effective permeability are
simultaneously negative attracts considerable attention of many
authors both experimentally and
theoretically\cite{Smith,Kong,Garcia,Jianqi,Shen}. In
1967\footnote{Note that, in the literature, some authors mentioned
the wrong year when Veselago suggested the {\it left-handed
media}. They claimed that Veselago proposed or introduced the
concept of {\it left-handed media} in 1968 or 1964. On the
contrary, the true history is as follows: Veselago's excellent
paper was first published in Russian in July, 1967 [Usp. Fiz. Nauk
{\bf 92}, 517-526 (1967)]. This original paper was translated into
English by W.H. Furry and published again in 1968 in the journal
of Sov. Phys. Usp.\cite{Veselago}. Unfortunately, Furry stated
erroneously in his English translation that the original version
of Veselago' work was first published in 1964.}, Veselago first
considered this peculiar medium and showed from Maxwellian
equations that such media having negative simultaneously negative
$\epsilon $ and $\mu $ exhibit a negative index of refraction,
{\it i.e.}, $n=-\sqrt{\epsilon \mu }$\cite{Veselago}. It follows
from the Maxwell's curl equations that the phase velocity of light
wave propagating inside this medium is pointed opposite to the
direction of energy flow, that is, the Poynting vector and wave
vector of electromagnetic wave would be antiparallel, {\it i.e.},
the vector {\bf {k}}, the electric field {\bf {E}} and the
magnetic field {\bf {H}} form a left-handed system; thus Veselago
referred to such materials as ``left-handed'' media, and
correspondingly, the ordinary medium in which {\bf {k}}, {\bf {E}}
and {\bf {H}} form a right-handed system may be termed the
``right-handed'' one. Other authors call this class of materials
``negative-index media (NIM)''\cite{Gerardin}, ``backward media
(BWM)''\cite{Lindell}, ``double negative media (DNM)'' and
Veselago's media. There exist a number of peculiar electromagnetic
and optical properties, for instance, many dramatically different
propagation characteristics stem from the sign change of the
optical refractive index and phase velocity, including reversal of
both the Doppler shift and Cerenkov radiation, anomalous
refraction, amplification of evanescent waves\cite{Pendryprl},
unusual photon tunneling\cite{Zhang}, modified spontaneous
emission rates and even reversals of radiation pressure to
radiation tension\cite{Smith}. In experiments, this artificial
negative electric permittivity media may be obtained by using the
{\it array of long metallic wires} (ALMWs)\cite{Pendry2}, which
simulates the plasma behavior at microwave frequencies, and the
artificial negative magnetic permeability media may be built up by
using small resonant metallic particles, {\it e.g.}, the {\it
split ring resonators} (SRRs), with very high magnetic
polarizability\cite{Pendry1}. A combination of the two structures
yields a left-handed medium. Recently, Shelby {\it et al.}
reported their first experimental realization of this artificial
composite medium, the permittivity and permeability of which have
negative real parts\cite{Smith}. One of the potential applications
of negative refractive index materials is to fabricate the
so-called ``superlenses'' (perfect lenses): specifically, a slab
of such materials may has the power to focus all Fourier
components of a 2D image, even those that do not propagate in a
radiative manner\cite{Pendryprl,Hooft}.

Engheta suggested that a slab of metamaterial with negative
electric permittivity and magnetic permeability (and hence
negative optical refractive index) can act as a phase
compensator/conjugator and, therefore, by combining such a slab
with another slab fabricated from a conventional (ordinary)
dielectric material one can, in principle, have a 1-D cavity
resonator whose dispersion relation may not depend on the sum of
thicknesses of the interior materials filling this cavity, but
instead it depends on the ratio of these thicknesses. Namely, one
can, in principle, conceptualize a 1-D compact, subwavelength,
thin cavity resonator with the total thickness far less than the
conventional $\frac{\lambda}{2}$\cite{Engheta}.

Engheta's idea for the 1-D compact, subwavelength, thin cavity
resonator is the two-layer rectangular structure (the left layer
of which is assumed to be a conventional lossless dielectric
material with permittivity and permeability being positive
numbers, and the right layer is taken to be a lossless
metamaterial with negative permittivity and permeability)
sandwiched between the two reflectors ({\it e.g.}, two perfectly
conducting plates)\cite{Engheta}. For the pattern of the 1-D
subwavelength cavity resonator readers may be referred to the
figures of reference\cite{Engheta}. Engheta showed that with the
appropriate choice of the ratio of the thicknesses $d_{1}$ to
$d_{2}$, the phase acquired by the incident wave at the left
(entrance) interface to be the same as the phase at the right
(exit) interface, essentially with no constraint on the total
thickness of the structure. The mechanism of this effect may be
understood as follows: as the planar electromagnetic wave exits
the first slab, it enters the rectangular slab of metamaterial and
finally it leaves this second slab. In the first slab, the
direction of the Poynting vector is parallel to that of phase
velocity, and in the second slab, however, these two vectors are
antiparallel with each other. Thus the wave vector $k_{2}$ is
therefore in the opposite direction of the wave vector $k_{1}$. So
the total phase difference between the front and back faces of
this two-layer rectangular structure is
$k_{1}d_{1}-|k_{2}|d_{2}$\cite{Engheta}. Therefore, whatever phase
difference is developed by traversing the first rectangular slab,
it can be decreased and even cancelled by traversing the second
slab. If the ratio of $d_{1}$ and $d_{2}$ is chosen to be
$\frac{d_{1}}{d_{2}}=\frac{|k_{2}|}{k_{1}}$, then the total phase
difference between the front and back faces of this two-layer
structure becomes zero ({\it i.e.}, the total phase difference is
not $2n\pi$, but instead of zero)\cite{Engheta}. As far as the
properties and phenomena in the subwavelength cavity resonators is
concerned, Tretyakov {\it et al.} investigated the evanescent
modes stored in cavity resonators with backward-wave
slabs\cite{Tretyakov}.

\section{A rectangular slab 1-D thin subwavelength cavity resonator}
To consider the 1-D wave propagation in a compact, subwavelength,
thin cavity resonator, we first take into account a slab cavity of
three-layer structure, where the regions 1 and 2 are located on
the left- and right- handed sides, and the plasmon-type medium (or
a superconductor material) is between the regions 1 and 2. The
above three-layer structure is assumed to be sandwiched between
the two reflectors (or two perfectly conducting
plates)\cite{Engheta}.  Assume that the wave vector of the
electromagnetic wave is parallel to the third component of
Cartesian coordinate. The electric and magnetic fields in the
region 1 (with the permittivity being $\epsilon_{1}$ and the
permeability being $\mu_{1}$) are written in the form
\begin{equation}
E_{x1}=E_{01}\sin \left(n_{1}k_{0}z\right),  \quad
H_{y1}=\frac{n_{1}k_{0}}{i\omega \mu_{1}}E_{01}\cos
\left(n_{1}k_{0}z\right)   \quad  (0\leq z\leq d_{1}), \label{eq1}
\end{equation}
where $k_{0}$ stands for the wave vector of the electromagnetic
wave under consideration in the free space, {\it i.e.},
$k_{0}=\frac{\omega}{c}$, and in the region 2, where $d_{1}+a\leq
z\leq d_{1}+d_{2}+a$, the electric and magnetic fields are of the
form
\begin{equation}
E_{x2}=E_{02}\sin
\left[n_{2}k_{0}\left(z-d_{1}-d_{2}-a\right)\right],   \quad
H_{y2}=\frac{n_{2}k_{0}}{i\omega \mu_{2}}E_{02}\cos
\left[n_{2}k_{0}\left(z-d_{1}-d_{2}-a\right)\right], \label{eq2}
\end{equation}
where $d_{1}$, $d_{2}$ and $a$ denote the thicknesses of the
regions 1, 2 and the plasmon (or superconducting) region,
respectively. The subscripts 1 and 2 in the present paper denote
the physical quantities in the regions 1 and 2. Note that here the
optical refractive indices $n_{1}$ and $n_{2}$ are defined to be
$n_{1}=\sqrt{\epsilon_{1}\mu_{1}}$ and
$n_{2}=\sqrt{\epsilon_{2}\mu_{2}}$. Although in the present paper
we will consider the wave propagation in the negative refractive
index media, the choice of the signs for $n_{1}$ and $n_{2}$ will
be irrelevant in the final results. So, we choose the plus signs
for $n_{1}$ and $n_{2}$ no matter whether the materials 1 and 2
are of left-handedness or not. The choice of the solutions
presented in (\ref{eq1}) and (\ref{eq2}) guarantees the
satisfaction of the boundary conditions at the perfectly
conducting plates at $z=0$ and $z=d_{1}+d_{2}+a$.

The electric and magnetic fields in the plasmon (or
superconducting) region (with the resonant frequency being
$\omega_{\rm p}$) take the form
\begin{equation}
E_{xs}=A\exp (\beta z)+B\exp (-\beta z),   \quad
H_{ys}=\frac{\beta}{i\omega}\left[A\exp (\beta z)-B\exp (-\beta
z)\right],                  \label{eq3}
\end{equation}
where the subscript $s$ represents the quantities in the plasmon
(or superconducting) region, and $\beta=\frac{\sqrt{\omega_{\rm
p}^{2}-\omega^{2}}}{c}$.

To satisfy the boundary conditions
\begin{equation}
E_{x1}|_{z=d_{1}}=E_{xs}|_{z=d_{1}},   \quad
H_{y1}|_{z=d_{1}}=H_{ys}|_{z=d_{1}}
\end{equation}
at the interface ($z=d_{1}$) between the region 1 and the plasmon
region, we should have
\begin{equation}
E_{01}\sin \left(n_{1}k_{0}d_{1}\right)=A\exp (\beta d_{1})+B\exp
(-\beta d_{1}),   \quad   \frac{n_{1}k_{0}}{\beta
\mu_{1}}E_{01}\cos \left(n_{1}k_{0}d_{1}\right)=A\exp (\beta
d_{1})-B\exp(-\beta d_{1}).                  \label{eq5}
\end{equation}
It follows that the parameters $A$ and $B$ in Eq.(\ref{eq3}) are
given as follows
\begin{eqnarray}
A&=&\frac{1}{2}\exp \left(-\beta d_{1}\right)E_{01}\left[\sin\left(n_{1}k_{0}d_{1}\right)+\frac{n_{1}k_{0}}{\beta \mu_{1}}\cos \left(n_{1}k_{0}d_{1}\right)\right],   \nonumber  \\                \nonumber \\
B&=&-\frac{1}{2}\exp \left(\beta
d_{1}\right)E_{01}\left[\frac{n_{1}k_{0}}{\beta \mu_{1}}\cos
\left(n_{1}k_{0}d_{1}\right)-\sin \left(n_{1}k_{0}d_{1}\right)
\right].                                               \label{eq6}
\end{eqnarray}
In the similar fashion, to satisfy the boundary conditions
\begin{equation}
E_{x2}|_{z=d_{1}+a}=E_{xs}|_{z=d_{1}+a},   \quad
H_{y2}|_{z=d_{1}+a}=H_{ys}|_{z=d_{1}+a}
\end{equation}
at the interface ($z=d_{1}+a$) between the region 2 and the
plasmon region, one should arrive at
\begin{eqnarray}
& & E_{02}\sin
\left(-n_{2}k_{0}d_{2}\right)=A\exp\left[\beta(d_{1}+a)\right]+B\exp\left[-\beta(d_{1}+a)\right],
\nonumber \\
& &    \frac{n_{2}k_{0}}{\beta \mu_{2}}E_{02}\cos
\left(-n_{2}k_{0}d_{2}\right)=A\exp\left[\beta(d_{1}+a)\right]-B\exp\left[-\beta(d_{1}+a)\right].
\label{eq7}
\end{eqnarray}
It follows that the parameters $A$ and $B$ in Eq.(\ref{eq3}) are
given as follows
\begin{eqnarray}
A&=&-\frac{1}{2}E_{02}\exp \left[-\beta (d_{1}+a)\right]\left[\sin\left(n_{2}k_{0}d_{2}\right)-\frac{n_{2}k_{0}}{\beta \mu_{2}}\cos \left(n_{2}k_{0}d_{2}\right)\right],   \nonumber  \\                \nonumber \\
B&=&-\frac{1}{2}\exp \left[-\beta
(d_{1}+a)\right]E_{02}\left[\frac{n_{2}k_{0}}{\beta \mu_{2}}\cos
\left(n_{2}k_{0}d_{2}\right)+\sin \left(n_{2}k_{0}d_{2}\right)
\right].                 \label{eq8}
\end{eqnarray}
Thus, according to Eq.(\ref{eq6}) and (\ref{eq8}), we can obtain
the following conditions
\begin{eqnarray}
& & E_{01}\left[\sin\left(n_{1}k_{0}d_{1}\right)+\frac{n_{1}k_{0}}{\beta \mu_{1}}\cos\left(n_{1}k_{0}d_{1}\right)\right]+E_{02}\exp(-\beta a)\left[\sin\left(n_{2}k_{0}d_{2}\right)-\frac{n_{2}k_{0}}{\beta \mu_{2}}\cos\left(n_{2}k_{0}d_{2}\right)\right]=0,   \nonumber  \\
                 \nonumber \\
& &  E_{01}\left[\frac{n_{1}k_{0}}{\beta
\mu_{1}}\cos\left(n_{1}k_{0}d_{1}\right)-\sin
\left(n_{1}k_{0}d_{1}\right)\right]-E_{02}\exp(\beta
a)\left[\frac{n_{2}k_{0}}{\beta \mu_{2}}\cos
\left(n_{2}k_{0}d_{2}\right)+\sin\left(n_{2}k_{0}d_{2}\right)\right]=0.
\label{eq10}
\end{eqnarray}
In order to have a nontrivial solution, {\it i.e.}, to have
$E_{01}\neq 0$ and $E_{02}\neq 0$, the determinant in
Eq.(\ref{eq10}) must vanish. Thus we obtain the following
restriction condition
\begin{eqnarray}
& &  \left[\exp(\beta a)+\exp(-\beta
a)\right]\left[\frac{n_{1}}{\mu_{1}}\tan
\left(n_{2}k_{0}d_{2}\right)+\frac{n_{2}}{\mu_{2}}\tan
\left(n_{1}k_{0}d_{1}\right)\right]
                                       \nonumber \\
& &   +\frac{\beta}{k_{0}}\left[\exp(\beta a)-\exp(-\beta
a)\right]\left[\tan \left(n_{1}k_{0}d_{1}\right)\tan
\left(n_{2}k_{0}d_{2}\right)+\frac{n_{1}n_{2}k_{0}^{2}}{\beta^{2}\mu_{1}\mu_{2}}\right]=0
\label{eq11}
\end{eqnarray}
for the electromagnetic wave in the three-layer-structure
rectangular cavity.

If the thickness, $a$, of the plasmon region is vanishing ({\it
i.e.}, there exists no plasmon region), then the restriction
equation (\ref{eq11}) is simplified to
\begin{equation}
\frac{n_{1}}{\mu_{1}}\tan\left(n_{2}k_{0}d_{2}\right)+\frac{n_{2}}{\mu_{2}}\tan
\left(n_{1}k_{0}d_{1}\right)=0.
\end{equation}
In what follows we will demonstrate why the introduction of
left-handed media will give rise to the novel design of the
compact thin cavity resonator. If the material in region 1 is a
regular medium while the one in region 2 is the left-handed
medium, it follows that
\begin{equation}
\frac{\tan\left(n_{1}k_{0}d_{1}\right)}{\tan\left(n_{2}k_{0}d_{2}\right)}=\frac{-n_{1}\mu_{2}}{n_{2}\mu_{1}}.
\end{equation}
According to Engheta\cite{Engheta}, this relation does not show
any constraint on the sum of thicknesses of $d_{1}$ and $d_{2}$.
It rather deals with the ratio of tangent of these thicknesses
(with multiplicative constants). If we assume that $\omega$,
$d_{1}$ and $d_{2}$ are chosen such that the small-argument
approximation can be used for the tangent function, the above
relation can be simplified as
\begin{equation}
\frac{d_{1}}{d_{2}}\simeq -\frac{\mu_{2}}{\mu_{1}}.
\end{equation}
This relation shows even more clearly how $d_{1}$ and $d_{2}$
should be related in order to have a nontrivial 1-D solution with
frequency $\omega$ for this cavity. So conceptually, what is
constrained here is $\frac{d_{1}}{d_{2}}$, not $d_{1}+d_{2}$.
Therefore, in principle, one can have a thin subwavelength cavity
resonator for a given frequency\cite{Engheta}.

 So, one of the most exciting ideas is the possibility to
design the so-called compact thin subwavelength cavity resonators.
It was shown that a pair of plane waves travelling in the system
of two planar slabs positioned between two metal planes can
satisfy the boundary conditions on the walls and on the interface
between two slabs even for arbitrarily thin layers, provided that
one of the slabs has negative material parameters\cite{Tretyakov}.

In the following let us take account of two interesting cases:

(i) If $\beta a\rightarrow 0$, then the restriction equation
(\ref{eq11}) is simplified to
\begin{equation}
\frac{n_{1}}{\mu_{1}}\tan
\left(n_{2}k_{0}d_{2}\right)+\frac{n_{2}}{\mu_{2}}\tan
\left(n_{1}k_{0}d_{1}\right)+\frac{2\beta^{2}a}{k_{0}}\left[\tan
\left(n_{1}k_{0}d_{1}\right)\tan
\left(n_{2}k_{0}d_{2}\right)+\frac{n_{1}n_{2}k_{0}^{2}}{\beta^{2}\mu_{1}\mu_{2}}\right]=0,
\end{equation}
which yields
\begin{equation}
\tan
\left(n_{1}k_{0}d_{1}\right)=\frac{k_{0}}{\beta}\frac{n_{1}}{\mu_{1}},
\quad           \tan
\left(n_{2}k_{0}d_{2}\right)=-\frac{k_{0}}{\beta}\frac{n_{2}}{\mu_{2}}.
\label{eq214}
\end{equation}
If both $n_{1}k_{0}d_{1}$ and $n_{2}k_{0}d_{2}$ are very small,
then one can arrive at $d_{1}\doteq \frac{1}{\beta \mu_{1}}$,
$d_{2}\doteq -\frac{1}{\beta \mu_{2}}$ from Eq.(\ref{eq214}),
which means that the thicknesses $d_{1}$ and $d_{2}$ depend upon
the plasmon parameter $\beta$.

(ii) If the resonant frequency $\omega_{\rm p}$ is very large (and
hence $\beta$), then it follows from Eq.(\ref{eq10}) and
(\ref{eq11}) that
\begin{equation}
\frac{n_{1}}{\mu_{1}}\tan \left(n_{2}k_{0}d_{2}\right)=0,   \quad
\frac{n_{2}}{\mu_{2}}\tan \left(n_{1}k_{0}d_{1}\right)=0,
\end{equation}
namely, regions 1 and 2 are isolated from each other., which is a
result familiar to us.

In conclusion, as was shown by Engheta, it is possible that when
one of the slab has a negative permeability, electromagnetic wave
in two adjacent slabs bounded by two metal walls can satisfy the
boundary conditions even if the distance between the two walls is
much smaller than the wavelength\cite{Engheta}.

\section{A cylindrical thin subwavelength cavity resonator}
Here we will consider the restriction equation for a cylindrical
 cavity to be a thin subwavelength cavity resonator containing left-handed media.
 It is well known that the Helmholtz equation $\nabla^{2}{\bf E}+k^{2}{\bf E}=0$
 in an axially symmetric cylindrical cavity (with the 2-D polar coordinates $\rho$ and $\varphi$) can be rewritten as
\begin{eqnarray}
& &  \nabla^{2}E_{\rho}-\frac{1}{\rho^{2}}E_{\rho}-\frac{2}{\rho^{2}}\frac{\partial E_{\varphi}}{\partial \varphi}+k^{2}E_{\rho}=0,                \nonumber \\
& &   \nabla^{2}E_{\varphi}-\frac{1}{\rho^{2}}E_{\varphi}+\frac{2}{\rho^{2}}\frac{\partial E_{\rho}}{\partial \varphi}+k^{2}E_{\varphi}=0,               \nonumber \\
& &  \nabla^{2}E_{z}+k^{2}E_{z}=0.       \label{eq201}
\end{eqnarray}
One can obtain the electromagnetic field distribution, $E_{\rho}$,
$E_{\varphi}$ and $H_{\rho}$, $H_{\varphi}$, in the above axially
symmetric cylindrical cavity via Eq.(\ref{eq201}). But here we
will adopt another alternative way to get the solutions of
electromagnetic fields in the cylindrical cavity. If the
electromagnetic fields are time-harmonic, {\it i.e.}, ${\bf
E}(\rho,\varphi,z,t)=\vec{{\mathcal
E}}(\rho,\varphi)\exp[i(hz-kct)]$ and ${\bf
H}(\rho,\varphi,z,t)=\vec{{\mathcal
H}}(\rho,\varphi)\exp[i(hz-kct)]$, then it follows from Maxwell
equations that
\begin{eqnarray}
& &  -ikc{\mathcal E}_{\rho}=\frac{1}{\epsilon}\left(\frac{1}{\rho}\frac{\partial {\mathcal H}_{z}}{\partial \varphi}-ih{\mathcal H}_{\varphi}\right),                \nonumber \\
& &  -ikc{\mathcal E}_{\varphi}=\frac{1}{\epsilon}\left(ih{\mathcal H}_{\rho}-\frac{\partial {\mathcal H}_{z}}{\partial \rho}\right),                \nonumber \\
& &  -ikc{\mathcal E}_{z}=\frac{1}{\epsilon}\left(\frac{\partial
{\mathcal H}_{\varphi}}{\partial \rho}+\frac{1}{\rho}{\mathcal
H}_{\varphi}-\frac{1}{\rho}\frac{\partial {\mathcal
H}_{\rho}}{\partial \varphi}\right),
\end{eqnarray}
and
\begin{eqnarray}
& &  ikc{\mathcal H}_{\rho}=\frac{1}{\mu}\left(\frac{1}{\rho}\frac{\partial {\mathcal E}_{z}}{\partial \varphi}-ih{\mathcal E}_{\varphi}\right),                \nonumber \\
& &  ikc{\mathcal H}_{\varphi}=\frac{1}{\mu}\left(ih{\mathcal E}_{\rho}-\frac{\partial {\mathcal E}_{z}}{\partial \rho}\right),                \nonumber \\
& &  ikc{\mathcal H}_{z}=\frac{1}{\mu}\left(\frac{\partial
{\mathcal E}_{\varphi}}{\partial \rho}+\frac{1}{\rho}{\mathcal
E}_{\varphi}-\frac{1}{\rho}\frac{\partial {\mathcal
E}_{\rho}}{\partial \varphi}\right).
\end{eqnarray}

Thus it is demonstrated that the electromagnetic fields ${\mathcal
E}_{\rho}$, ${\mathcal E}_{\varphi}$ and ${\mathcal H}_{\rho}$,
${\mathcal H}_{\varphi}$ can be expressed in terms of ${\mathcal
E}_{z}$ and ${\mathcal H}_{z}$, {\it i.e.},
\begin{equation}
{\mathcal E}_{\rho}=\frac{i}{k^{2}-h^{2}}\left(h\frac{\partial
{\mathcal E}_{z}}{\partial \rho}+\frac{k^{2}}{\epsilon
\rho}\frac{\partial {\mathcal H}_{z}}{\partial \varphi}\right),
\quad {\mathcal
E}_{\varphi}=\frac{i}{k^{2}-h^{2}}\left(h\frac{1}{\rho}\frac{\partial
{\mathcal E}_{z}}{\partial
\varphi}-\frac{k^{2}}{\epsilon}\frac{\partial {\mathcal
H}_{z}}{\partial \rho}\right), \label{eq2036}
\end{equation}
and
\begin{equation}
{\mathcal H}_{\rho}=\frac{i}{k^{2}-h^{2}}\left(h\frac{\partial
{\mathcal H}_{z}}{\partial \rho}-\frac{k^{2}}{\mu
\rho}\frac{\partial {\mathcal E}_{z}}{\partial \varphi}\right),
\quad {\mathcal
H}_{\varphi}=\frac{i}{k^{2}-h^{2}}\left(h\frac{1}{\rho}\frac{\partial
{\mathcal H}_{z}}{\partial
\varphi}+\frac{k^{2}}{\mu}\frac{\partial {\mathcal
E}_{z}}{\partial \rho}\right).      \label{eq2037}
\end{equation}
As an illustrative example, in what follows, we will consider only
the TM wave ({\it i.e.}, ${\mathcal H}_{z}=0$) in the axially
symmetric double-layer cylindrical thin subwavelength cavity
resonator. Assume that the permittivity, permeability and radius
of media in the interior and exterior layers of this double-layer
cavity resonator are $\epsilon_{1}$, $\mu_{1}$, $R_{1}$ and
$\epsilon_{2}$, $\mu_{2}$, $R_{2}$, respectively. According to the
Helmholtz equation (with the boundary material being the perfectly
conducting medium, ${\mathcal E}_{2z}|_{R_{2}}=0$), we can obtain
${\it
E}_{1z}=J_{m}\left(\sqrt{k_{1}^{2}-h_{1}^{2}}\rho\right)\left\{
{\begin{array}{*{20}c}
   {\cos m\varphi }  \\
   {\sin m\varphi}  \\
\end{array}}\right\}$ and ${\it
E}_{2z}=\left[AJ_{m}\left(\sqrt{k_{2}^{2}-h_{2}^{2}}\rho\right)+BN_{m}\left(\sqrt{k_{2}^{2}-h_{2}^{2}}\rho\right)\right]\left\{
{\begin{array}{*{20}c}
   {\cos m\varphi }  \\
   {\sin m\varphi}  \\
\end{array}}\right\}$. Thus it follows from Eq.(\ref{eq2036}) and
(\ref{eq2037}) that the electromagnetic fields in both interior
and exterior layers are of the form
\begin{eqnarray}
{\it
E}_{1\rho}&=&\frac{ih_{1}}{\sqrt{k_{1}^{2}-h_{1}^{2}}}J'_{m}\left(\sqrt{k_{1}^{2}-h_{1}^{2}}\rho\right)\left\{
{\begin{array}{*{20}c}
   {\cos m\varphi }  \\
   {\sin m\varphi}  \\
\end{array}}\right\},                \nonumber \\
{\it
E}_{1\varphi}&=&\frac{imh_{1}}{\left(k_{1}^{2}-h_{1}^{2}\right)\rho}J_{m}\left(\sqrt{k_{1}^{2}-h_{1}^{2}}\rho\right)\left\{
{\begin{array}{*{20}c}
   {\sin m\varphi }  \\
   {-\cos m\varphi}  \\
\end{array}}\right\},
                          \nonumber \\
   {\it
E}_{1z}&=&J_{m}\left(\sqrt{k_{1}^{2}-h_{1}^{2}}\rho\right)\left\{
{\begin{array}{*{20}c}
   {\cos m\varphi }  \\
   {\sin m\varphi}  \\
\end{array}}\right\},
\end{eqnarray}

\begin{eqnarray}
{\mathcal
H}_{1\rho}&=&\frac{imk_{1}^{2}}{\left(k_{1}^{2}-h_{1}^{2}\right)\rho}\frac{1}{\mu_{1}}J_{m}\left(\sqrt{k_{1}^{2}-h_{1}^{2}}\rho\right)\left\{
{\begin{array}{*{20}c}
   {-\sin m\varphi }  \\
   {\cos m\varphi}  \\
\end{array}}\right\},                  \nonumber \\
{\mathcal
H}_{1\varphi}&=&\frac{ik_{1}^{2}}{\sqrt{k_{1}^{2}-h_{1}^{2}}}\frac{1}{\mu_{1}}J'_{m}\left(\sqrt{k_{1}^{2}-h_{1}^{2}}\rho\right)\left\{
{\begin{array}{*{20}c}
   {\cos m\varphi }  \\
   {\sin m\varphi}  \\
\end{array}}\right\}, \nonumber \\
{\mathcal H}_{1z}&=& 0,
\end{eqnarray}
and
\begin{eqnarray}
{\it
E}_{2\rho}&=&\frac{ih_{2}}{\sqrt{k_{2}^{2}-h_{2}^{2}}}\left[AJ'_{m}\left(\sqrt{k_{2}^{2}-h_{2}^{2}}\rho\right)+BN'_{m}\left(\sqrt{k_{2}^{2}-h_{2}^{2}}\rho\right)\right]\left\{
{\begin{array}{*{20}c}
   {\cos m\varphi }  \\
   {\sin m\varphi}  \\
\end{array}}\right\},                \nonumber \\
{\it
E}_{2\varphi}&=&\frac{imh_{2}}{\left(k_{2}^{2}-h_{2}^{2}\right)\rho}\left[AJ_{m}\left(\sqrt{k_{2}^{2}-h_{2}^{2}}\rho\right)+BN_{m}\left(\sqrt{k_{2}^{2}-h_{2}^{2}}\rho\right)\right]\left\{
{\begin{array}{*{20}c}
   {\sin m\varphi }  \\
   {-\cos m\varphi}  \\
\end{array}}\right\},
                          \nonumber \\
   {\it
E}_{2z}&=&\left[AJ_{m}\left(\sqrt{k_{2}^{2}-h_{2}^{2}}\rho\right)+BN_{m}\left(\sqrt{k_{2}^{2}-h_{2}^{2}}\rho\right)\right]\left\{
{\begin{array}{*{20}c}
   {\cos m\varphi }  \\
   {\sin m\varphi}  \\
\end{array}}\right\},
\end{eqnarray}

\begin{eqnarray}
{\mathcal
H}_{2\rho}&=&\frac{imk_{2}^{2}}{\left(k_{2}^{2}-h_{2}^{2}\right)\rho}\frac{1}{\mu_{2}}\left[AJ_{m}\left(\sqrt{k_{2}^{2}-h_{2}^{2}}\rho\right)+BN_{m}\left(\sqrt{k_{2}^{2}-h_{2}^{2}}\rho\right)\right]\left\{
{\begin{array}{*{20}c}
   {-\sin m\varphi }  \\
   {\cos m\varphi}  \\
\end{array}}\right\},                  \nonumber \\
{\mathcal
H}_{2\varphi}&=&\frac{ik_{2}^{2}}{\sqrt{k_{2}^{2}-h_{2}^{2}}}\frac{1}{\mu_{2}}\left[AJ'_{m}\left(\sqrt{k_{2}^{2}-h_{2}^{2}}\rho\right)+BN'_{m}\left(\sqrt{k_{2}^{2}-h_{2}^{2}}\rho\right)\right]\left\{
{\begin{array}{*{20}c}
   {\cos m\varphi }  \\
   {\sin m\varphi}  \\
\end{array}}\right\}, \nonumber \\
{\mathcal H}_{2z}&=& 0,
\end{eqnarray}
where $J_{m}$ and $N_{m}$ denote Bessel functions and Norman
functions, respectively, and
$k_{1}=\sqrt{\epsilon_{1}\mu_{1}}\frac{\omega}{c}$,
$k_{2}=\sqrt{\epsilon_{2}\mu_{2}}\frac{\omega}{c}$.

By using the boundary conditions ${\mathcal
E}_{2z}|_{\rho=R_{2}}=0$, ${\mathcal
E}_{2\varphi}|_{\rho=R_{2}}=0$ (due to the perfectly conducting
medium at the boundary $\rho=R_{2}$), one can determine the
relationship between $A$ and $B$, {\it i.e.},
\begin{equation}
 B=-A\frac{J_{m}\left(\sqrt{k_{2}^{2}-h_{2}^{2}}R_{2}\right)}{N_{m}\left(\sqrt{k_{2}^{2}-h_{2}^{2}}R_{2}\right)}.
 \label{eq2010}
\end{equation}
By using the boundary conditions ${\mathcal
E}_{1z}|_{R1}={\mathcal E}_{2z}|_{R1}$, ${\mathcal
E}_{1\varphi}|_{R1}={\mathcal E}_{2\varphi}|_{R1}$, one can obtain
\begin{eqnarray}
& & J_{m}\left(\sqrt{k_{1}^{2}-h_{1}^{2}}R_{1}\right)=AJ_{m}\left(\sqrt{k_{2}^{2}-h_{2}^{2}}R_{1}\right)+BN_{m}\left(\sqrt{k_{2}^{2}-h_{2}^{2}}R_{1}\right),                 \nonumber \\
& &
\frac{h_{1}}{k_{1}^{2}-h_{1}^{2}}J_{m}\left(\sqrt{k_{1}^{2}-h_{1}^{2}}R_{1}\right)=\frac{h_{2}}{k_{2}^{2}-h_{2}^{2}}\left[AJ_{m}\left(\sqrt{k_{2}^{2}-h_{2}^{2}}R_{1}\right)+BN_{m}\left(\sqrt{k_{2}^{2}-h_{2}^{2}}R_{1}\right)\right].
\label{eq2011}
\end{eqnarray}
The second equation in Eq.(\ref{eq2011}) is employed to determine
the relation between $h_{1}$ and $h_{2}$. With the help of the
boundary conditions ${\mathcal H}_{1z}|_{R1}={\mathcal
H}_{2z}|_{R1}$ , ${\mathcal H}_{1\varphi}|_{R1}={\mathcal
H}_{2\varphi}|_{R1}$, one can arrive at
\begin{equation}
\frac{n_{1}}{\sqrt{k_{1}^{2}-h_{1}^{2}}}\frac{1}{\mu_{1}}J'_{m}\left(\sqrt{k_{1}^{2}-h_{1}^{2}}R_{1}\right)=\frac{n_{2}}{\sqrt{k_{2}^{2}-h_{2}^{2}}}\frac{1}{\mu_{2}}\left[AJ'_{m}\left(\sqrt{k_{2}^{2}-h_{2}^{2}}R_{1}\right)+BN'_{m}\left(\sqrt{k_{2}^{2}-h_{2}^{2}}R_{1}\right)\right].
\label{eq2012}
\end{equation}
Combination of the first equation in Eq.(\ref{eq2011}) and
(\ref{eq2012}), we have
\begin{equation}
\frac{n_{1}}{\mu_{1}\sqrt{k_{1}^{2}-h_{1}^{2}}}\frac{J'_{m}\left(\sqrt{k_{1}^{2}-h_{1}^{2}}R_{1}\right)}{J_{m}\left(\sqrt{k_{1}^{2}-h_{1}^{2}}R_{1}\right)}=\frac{n_{2}}{\mu_{2}\sqrt{k_{2}^{2}-h_{2}^{2}}}\frac{AJ'_{m}\left(\sqrt{k_{2}^{2}-h_{2}^{2}}R_{1}\right)+BN'_{m}\left(\sqrt{k_{2}^{2}-h_{2}^{2}}R_{1}\right)}{AJ_{m}\left(\sqrt{k_{2}^{2}-h_{2}^{2}}R_{1}\right)+BN_{m}\left(\sqrt{k_{2}^{2}-h_{2}^{2}}R_{1}\right)},
\end{equation}
which may be viewed as the restriction condition for the
cylindrical cavity resonator. If we choose a typical case with
$h_{1}=h_{2}=0$, then the obtained restriction condition is
simplified to be
\begin{equation}
\frac{1}{\mu_{1}}\frac{J'_{m}\left(k_{1}R_{1}\right)}{J_{m}\left(k_{1}R_{1}\right)}=\frac{1}{\mu_{2}}\frac{AJ'_{m}\left(k_{2}R_{1}\right)+BN'_{m}\left(k_{2}R_{1}\right)}{AJ_{m}\left(k_{2}R_{1}\right)+BN_{m}\left(k_{2}R_{1}\right)}.
\label{2013}
\end{equation}
Substitution of the relation (\ref{eq2010}) into (\ref{2013})
yields
\begin{equation}
\frac{1}{\mu_{1}}\frac{J'_{m}\left(k_{1}R_{1}\right)}{J_{m}\left(k_{1}R_{1}\right)}=\frac{1}{\mu_{2}}\frac{J'_{m}\left(k_{2}R_{1}\right)N_{m}\left(k_{2}R_{2}\right)-J_{m}\left(k_{2}R_{2}\right)N'_{m}\left(k_{2}R_{1}\right)}{J_{m}\left(k_{2}R_{1}\right)N_{m}\left(k_{2}R_{2}\right)-J_{m}\left(k_{2}R_{2}\right)N_{m}\left(k_{2}R_{1}\right)}.
\label{eq2015}
\end{equation}
Eq.(\ref{eq2015}) is just the simplified restriction condition for
the cylindrical cavity resonator.

Similar to the analysis presented in Sec. I, it is readily shown
that by introducing the left-handed media such cylindrical cavity
will also act as a compact thin subwavelength resonator. The
discussion on this subject will not be performed further in the
present paper.
\section{A spherical thin subwavelength cavity resonator}
Here we will consider briefly the restriction equation for a
spherical cavity to be a thin subwavelength cavity resonator
containing left-handed media. Assume that the permittivity,
permeability and radius of media in the interior and exterior
layers of this double-layer cavity resonator are $\epsilon_{1}$,
$\mu_{1}$, $\rho_{1}$ and $\epsilon_{2}$, $\mu_{2}$, $\rho_{2}$,
respectively, and that the boundary medium at $\rho=\rho_{2}$ is
the perfectly conducting material. Note that here the functions,
symbols and quantities are adopted in the paper\cite{Shen}, {\it
e.g.}, $N_{1}$, $N_{2}$ denote the refractive indices of interio
and exterior layers, and $j_{n}$ and $h_{n}^{(1)}$ are the
spherical Bessel and Hankel functions.

According to the paper\cite{Shen}, it follows from the boundary
condition at $\rho=\rho_{2}$ that

\begin{eqnarray}
& & a^{m}_{n}j_{n}(N_{2}\rho_{2})+a_{n}^{\bar{m}}h^{(1)}_{n}(N_{2}\rho_{2})=0,                 \nonumber \\
& &
b^{m}_{n}[N_{2}\rho_{2}j_{n}(N_{2}\rho_{2})]'+b_{n}^{\bar{m}}\left[N_{2}\rho_{2}h^{(1)}_{n}(N_{2}\rho_{2})\right]'=0.
\label{eq41}
\end{eqnarray}
The roles of Eqs.(\ref{eq41}) is to determine the Mie coefficients
$a_{n}^{\bar{m}}$ and $b_{n}^{\bar{m}}$ in terms of $a^{m}_{n}$
and $b^{m}_{n}$.

At the boundary $\rho=\rho_{1}$, it follows from the boundary
condition ${\bf i}_{1}\times {\bf E}_{\rm m}={\bf i}_{1}\times
{\bf E}_{\rm t}$ that one can obtain
\begin{eqnarray}
& &   a^{m}_{n}j_{n}(N_{2}\rho_{1})+a_{n}^{\bar{m}}h^{(1)}_{n}(N_{2}\rho_{1})=a_{n}^{t}j_{n}(N_{1}\rho_{1}),               \nonumber \\
& &
N_{1}b^{m}_{n}\left[N_{2}\rho_{1}j_{n}(N_{2}\rho_{1})\right]'+N_{1}b_{n}^{\bar{m}}\left[N_{2}\rho_{1}h^{(1)}_{n}(N_{2}\rho_{1})\right]'=N_{2}b_{n}^{t}\left[N_{1}\rho_{1}j_{n}(N_{1}\rho_{1})\right]'.
\label{eq42}
\end{eqnarray}
The role of the first and second equations in Eq.(\ref{eq42}) is
to obtain the expressions for the Mie coefficients $a_{n}^{t}$ and
$b_{n}^{t}$ in terms of $a^{m}_{n}$ and $b^{m}_{n}$, respectively.

In the same manner, at the boundary $\rho=\rho_{1}$, it follows
from the boundary condition ${\bf i}_{1}\times {\bf H}_{\rm
m}={\bf i}_{1}\times {\bf H}_{\rm t}$ that one can obtain
\begin{eqnarray}
& &   \mu_{1}\left\{a^{m}_{n}\left[N_{2}\rho_{1}j_{n}(N_{2}\rho_{1})\right]'+a_{n}^{\bar{m}}\left[N_{2}\rho_{1}h^{(1)}_{n}(N_{2}\rho_{1})\right]'\right\}=\mu_{2}a_{n}^{t}\left[N_{1}\rho_{1}j_{n}(N_{1}\rho_{1})\right]',               \nonumber \\
& &    N_{2}
\mu_{1}\left[b^{m}_{n}j_{n}(N_{2}\rho_{1})+b_{n}^{\bar{m}}h^{(1)}_{n}(N_{2}\rho_{1})\right]=N_{1}\mu_{2}b_{n}^{t}j_{n}(N_{1}\rho_{1}).
\label{eq43}
\end{eqnarray}
Insertion of the expressions for the Mie coefficients
$a_{n}^{\bar{m}}$ and $b_{n}^{\bar{m}}$ and $a_{n}^{t}$ and
$b_{n}^{t}$ in terms of $a^{m}_{n}$ and $b^{m}_{n}$ obtained by
Eqs.(\ref{eq41}) and (\ref{eq42}) into Eqs.(\ref{eq43}) will lead
to a set of equations of the Mie coefficients $a^{m}_{n}$ and
$b^{m}_{n}$. In order to have a nontrivial solutions of
$a^{m}_{n}$ and $b^{m}_{n}$, the determinant in Eqs.(\ref{eq43})
must vanish. Thus we will obtain a restriction condition for the
spherical cavity resonator.

By analogy with the analysis presented in Sec. I, it is easily
verified that by involving the left-handed media such spherical
cavity will also serve as a compact thin subwavelength resonator.
In a word, the possibility to satisfy the boundary conditions for
small distances between metal plates is based on the fact that
plane waves in Veselago media are backward waves, meaning that the
phase shift due to propagation in a usual slab can be compensated
by a negative phase shift inside a backward-wave
slab\cite{Engheta,Tretyakov}. We will not discuss further this
topic in this paper.

\textbf{Acknowledgements}
 The author was grateful to the National
Natural Science Foundation of China (NNSFC) for their support by
NNSFC Research Grant No. $90101024$ and $60378037$.

\end{document}